# XCAT3.0: A Comprehensive Library of Personalized Digital Twins Derived from CT scans

Lavsen Dahal[a,b], Mobina Ghojoghnejad[a], Liesbeth Vancoillie[a],

Dhrubajyoti Ghosh[a,c], Yubraj Bhandari[a], David Kim[a,b], Fong Chi Ho[a,b],

Fakrul Islam Tushar[a,b], Sheng Luo[a,c], Kyle J. Lafata[a,b,d], Ehsan Abadi[a,b],

Ehsan Samei[a,b], Joseph Y. Lo[a,b,+], W. Paul Segars[a,+]

[a]*Center for Virtual Imaging Trials, Carl E. Ravin Advanced Imaging Laboratories, Department of Radiology, Duke University School of Medicine, Durham, NC, 27708, USA*

[b]*Electrical and Computer Engineering, Pratt School of Engineering, Duke University, Durham, NC, 27708, USA*

[c]*Department of Biostatistics & Bioinformatics, Duke University School of Medicine, Durham, NC, 27708, USA*

[d]*Department of Radiation Oncology, Duke University School of Medicine, Durham, NC, 27708, USA*

+These two authors contributed equally as Co-Senior Authors

## Abstract

Virtual Imaging Trials (VIT) offer a cost-effective and scalable approach for evaluating medical imaging technologies. Computational phantoms, which mimic real patient anatomy and physiology, play a central role in VITs. However, the current libraries of computational phantoms face limitations, particularly in terms of sample size and diversity. Insufficient representation of the population hampers accurate assessment of imaging technologies across different patient groups. Traditionally, the more realistic computational phantoms were created by manual segmentation, which is a laborious and time-consuming task, impeding the expansion of phantom libraries. This study presents a framework for creating realistic computational phantoms using a suite of automatic segmentation models and performing three forms of automated quality control on the segmented organ masks. The result is the release of over 2500 new computational phantoms, so-named XCAT

3.0 after the ubiquitous XCAT computational construct. This new formation embodies 140 structures and represents a comprehensive approach to detailed anatomical modeling. The developed computational phantoms are formatted in both voxelized and surface mesh formats. The framework is combined with an in-house CT scanner simulator to produce realistic CT images. The framework has the potential to advance virtual imaging trials, facilitating comprehensive and reliable evaluations of medical imaging technologies. Phantoms may be requested at https://cvit.duke.edu/resources/. Code, model weights, and sample CT images are available at https://xcat-3.github.io/.

# 1 Introduction

Compared to traditional trials, virtual imaging trials (VIT) can be used to evaluate and optimize medical imaging technologies in ways that are cost effective, fast, and scalable [1-4]. VITs require a virtual patient library that mimics a clinical population in their anatomy and physiology. Such a virtual patient dataset can be created using computational anthropomorphic phantoms. These simulated phantoms are expected to represent the true population in terms of relevant trial characteristics such as body habitus, organ volumes, tissue properties, and blood flow.

There have been considerable efforts toward developing computational phantoms from geometric models in the 1960s [5] to more advanced, surface-based models in recent years [6]. Although many different phantoms have been developed over the years [7, 8], the current phantom library sample size [9-12] is inadequate for use in large-scale virtual imaging studies where the required sample size is estimated to be in the thousands [13-15]. It is challenging to generate such a large number of computational phantoms because realistic phantom developments have, up to now, relied on manual organ segmentation, which is painstakingly slow, requires skilled expertise, and is still prone to human variability. Additionally, after segmentation, phantoms undergo multiple post-processing steps to ensure precise 3D representation as surface models for various downstream tasks. This process is susceptible to errors and often necessitates significant manual intervention.

Recent advancements in deep learning have improved organ segmentations [16-20], enabling the creation of patient specific phantoms with accurate organ geometry representation. Leveraging that approach, a recent study [21] aimed to automate phantom generation from radiological images but was limited by the number of organs segmented and lack of a quality assurance process. One segmentation algorithm which has gained popularity is TotalSegmentator v2 [16] which can identity 117 anatomical structures given 3D CT imaging data. The Segment Anything Model [19] is another such algorithm which excels in 2D segmentation [22], however is limited in medical imaging segmentation [22]. Despite rapid advancements in the field, none of these models provided a scalable solution to address the issues of segmentation failures or implement comprehensive quality control.

In this study, we aimed to address the prior limitation towards building a large robust set of unique computational phantoms. Specifically, we devised and released a deep learning segmentation model (DukeSeg) capable of segmenting up to 140 structures integrated with an automated quality control in three distinct ways to effectively address segmentation failures. This effort was then used to curate over 2500 anatomical models in both voxel and high-resolution mesh output formats, each derived from actual patient data. These models are enriched with crucial metadata drawn from the original patients such as race, sex, age, and body habitus. This personalized dual-format approach is crucial for detailed anatomical studies, virtual imaging trials, surgical planning, and virtual reality simulations, allowing for smoother visualizations and more precise model interactions. The generation of models followed a pipeline that integrated and automated multiple complex steps: segmentation, quality control, meshing, voxelization, and enhancement, as illustrated in Figure 1.

## 2 Methods

*2.1 Data Curation*

Four datasets with a total of 3400 CT volumes were used to develop the segmentation model, as presented in Table 1 and Table 2. The first set of data included the public dataset associated with TotalSegmentator [16] and a private dataset from Duke University Health System, which were used for model development and validation, while four other datasets were used for external testing: CT-ORG [23], AbdomenCT-1K [24], AMOS [25] and XCAT [10, 11]. The segmentation model was applied to a separate dataset of 3581 CT volumes from Duke University Health System. The inclusion criteria for the dataset consisted of CT volumes acquired between January 1, 2014, and March 30, 2023, utilizing protocols that include PET/CT and chest-abdomen-pelvis scans, ensuring a comprehensive coverage of the whole body.

Our segmentation model training drew upon public and private data that were pseudo-labeled using a combination of public and private segmentation models. This approach allowed us to expand the training data and to incorporate a wider range of anatomical variations and complexities. The training data of 990 patient CT volumes comprised 489 patients selected from the 600 in the TotalSegmentator [16] dataset that contained full-body structures and 501 from the Duke CT dataset which were randomly sampled. For the TotalSegmentator cases, the provided labels were used. For the Duke data, pseudo-labels of organ masks were generated automatically using a combination of three models. The TotalSegmentator model was used to label 104 structures while a proprietary commercial model (an early version of organ segmentation tool by Siemens Healthineers) was used to label 33 structures. Additionally, the MOOSE public body composition model [20] created labels for visceral fat, subcutaneous fat, and muscles.

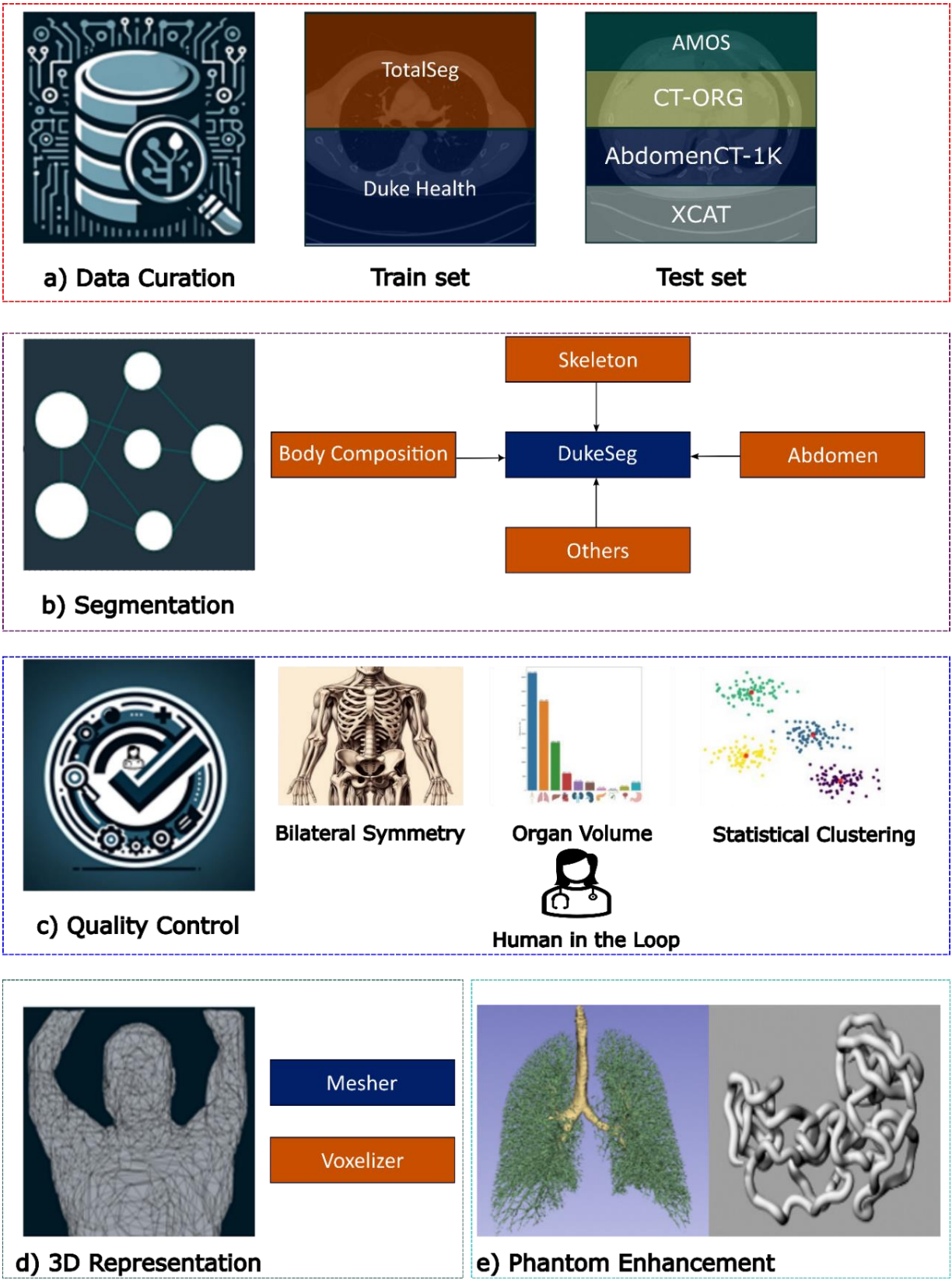

*Figure 1 Overview of the Methodology: This figure illustrates the five-module approach used in our study. (a) Data Curation: Collect data from various public and private repositories to train and test the deep learning-based segmentation model, DukeSeg. (b) DukeSeg Model: An in-house developed segmentation tool consisting of four distinct models tailored for segmenting different anatomical structures. (c) Quality Control: Utilizes four types of quality assessments, validated by the physician to ensure accuracy through a human-in-the-loop approach. (d) 3D Representation: Represents the 3D segmentation results in mesh or voxel formats. (e) Detailed Modeling: Includes mathematical modeling of intricate anatomical details such as the small intestine, and airways and vessels.*

| Table 1. Split for Train and Test utilizing various public and private datasets. | | |
|---|---|---|
| *Dataset* | *#Train* | *#Test* |
| TotalSegmentator | 489 | - |
| DukeCT | 501 | - |
| CT-ORG | - | 140 |
| AbdomenCT-1K | - | 1000 |
| AMOS | - | 200 |
| XCAT | - | 50 |
| Total | 990 | 1420 |

| Table 2. Characteristics of Training set | | |
|---|---|---|
| Dataset | Totalseg | DukeCT |
| #Patients | 489 | 501 |
| Male | 288 | 354 |
| Female | 201 | 147 |
| Age (years) | *63.5 (14.9)* | *62.1 (16.3)* |
| Weight (kg) | - | *86.5 (23.1)* |
| Height (m) | - | *1.73 (0.10)* |
| Slice thickness (mm) | *1.5* | *0.625, 3.75* |

## 2.2 Deep Learning Based Segmentation

The nnU-Net deep learning-based method [26] was trained as a set of four models, each specializing in segmenting different sets of structures. The first model segmented the body outline, subcutaneous fat, visceral fat, and muscles. The second model focused exclusively on bone segmentation, encompassing a set of 62 classes. The third model segmented a total of 13 classes focusing on the internal organs in the abdomen region and the esophagus. The final model segmented 61 additional structures. In total, the output labels of each of four models were deployed to offer the highest segmentation coverage of a diverse range of anatomical structures.

## 2.3 Quality Control

With any segmentation approach, failures can occur due to factors like poor image quality or limitations of the model. Initially, we excluded all the patients with age less than 14 years old as the segmentation algorithm was not effective for younger pediatric patients. Subsequently, we deployed four approaches for the quality control module: bilateral symmetry, volume thresholding, statistical outlier detection, and human screening. First, by assuming bilateral symmetry in the skeletal system, differences in volume exceeding 50% were used to indicate segmentation issues for symmetrical bones such as ribs, clavicles, pelvis, and femurs. Up to 2 discrepancies among the 16 sets of symmetrical bones were evaluated, which included 12 pairs of ribs and other paired skeletal structures. The bones of the arms were excluded due to position variations during acquisition. Second, the organ volume was computed after each segmentation and used to flag patients with more than 25% of segmented structures having zero volume, which indicated unusual positioning or incorrect scanning. Third, in the statistical outlier detection method, Hartigan's dip test was applied to determine if the volume distribution is multimodal. For organs based on several conditions below, unimodal outliers were identified by the interquartile range and multimodal outliers by Gaussian mixture models. The scans with incomplete or missing skull were identified based on outlier probabilities for three structures: brainstem, brain, and skull. Additionally, patients were flagged where more than two organs exceeded an outlier probability threshold of 0.9. For patients with multiple CT volumes, the scan with the lowest average outlier probability was selected such that the library represented only unique patients. In our final quality control step, segmentation volumes rendered as 2D images were examined by a physician who flagged any observed anomalies. This methodology exemplifies a 'human-in-the-loop' approach, combining expert oversight with technical optimizations for expedited yet thorough analysis. Together, the above quality control steps would reject cases if they exhibited major faults, such as failures in bilateral symmetry, partial or incomplete scans as evidenced by volume checks, failure in segmenting at least two organs identified by statistical clustering method, or human screening of rendered volumes.

*2.4 3D Representation & Phantoms Smoothing*

Cases that passed the above quality control criteria proceeded to the 3D representation module, where voxelized segmentation masks were transformed into smooth polygon meshes. For cases with minor irregularities, such as fewer segmented organs, these details were documented in the metadata along with a quality rating from a medical doctor.

For mesh transformation, GPU-accelerated Laplacian smoothing was used to iteratively refine the vertex positions of 3D meshes while preserving their overall shapes. By utilizing a sparse adjacency matrix to represent mesh connectivity and incorporating a customizable smoothing weight, the degree of smoothing could be fine-tuned. The algorithm began by converting input mesh data into tensors. It then constructed edges, initialized edge weights, and created a sparse adjacency matrix. Subsequently, row sums were computed for normalization, and the main smoothing loop iteratively updated vertex positions to return phantom mesh representations that were faithfully refined.

We developed a web application that showcases the developed phantoms interactively. This user-friendly application allows exploration of phantoms filtered by age, sex, and race, Body Mass Index (BMI), with options to select and examine specific structures as shown in Figure 2.

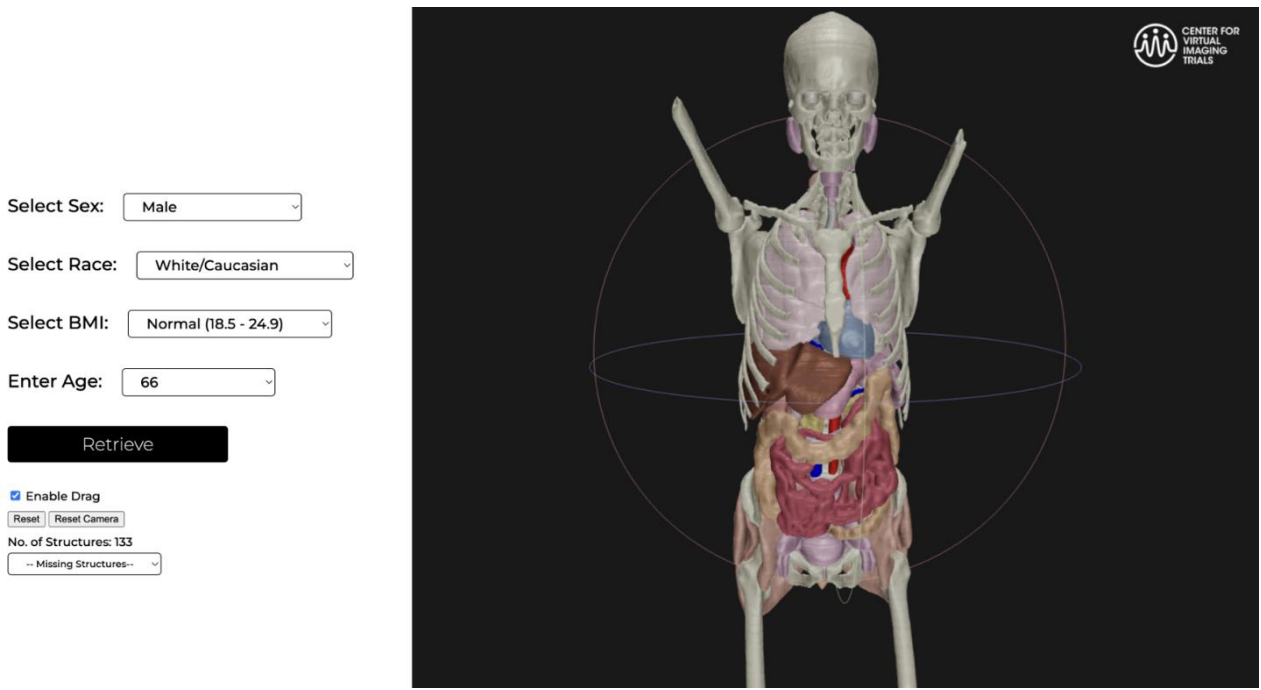

*Figure 2. Web application that shows the 3d anatomical models while also allowing user interaction. The application can be accessed at https://xcat-3.github.io.*

## 2.5 Phantom enhancements

The basic phantoms consist of binary masks of the organs and structures, which can be optionally enhanced using techniques developed in related studies from our laboratory. Cardiac and respiratory motion can be incorporated into the phantoms using the same techniques for the XCAT series of adult and pediatric phantoms [10, 11]. Airways and vessels within the lungs can be procedurally extended from the initial segmented branches toward terminal branches, thus distributing flow through the lungs while avoiding intersections between structures [27]. Texture for lung parenchyma can be added to voxelized phantoms for increased anatomical detail [28]. Similarly, bones can be enhanced to generate the inner trabecular bone structure with geometric and topological properties indicative of the individual bones with normal and abnormal variations [29]. The developed phantoms can also incorporate models of disease, including lung lesions [30, 31], cardiac plaque [32, 33], emphysema [34], bronchitis [35], COVID-19 pneumonia [36], and liver lesions [31].

Organ masks can also be refined through synthesis or generative procedures, which can be advantageous for certain challenging organs. For the small intestines, the segmentation mask can be used to guide the procedural generation of a contiguous, tubular surface with length and diameter within the range of normal values [37] [38]. The large intestine segmentations can also be refined using a combination of geometric deep learning and denoising diffusion models [39]. The result is a more realistic and flexible model for the intestines, which can be manipulated to simulate respiratory or gastrointestinal motion [40].

*2.6 Imaging Simulation using the Phantoms*

To generate simulated medical imaging scans, the phantoms produced through our pipeline may be input into imaging simulators [41-44] . One example is our in-house CT simulation platform (DukeSim) which has a hybrid approach for simulating CT projections using ray-tracing and Monte Carlo modules [44, 45]. DukeSim simulates a broad range of physical elements of the image acquisition process factors including different CT scanner designs and reconstruction algorithms. This simulation process culminates in the generation of highly detailed and realistic simulated CT images. Completing the virtual imaging chain, the simulated images can then undergo final analysis such as by observer models or deep learning algorithms.

# 3 Results

3.1 Segmentation

The box plots for segmentation in Figure 3 indicates that our model performed similar to the other two models as evaluated using Dice similarity coefficient (DSC) between the predicted mask and the reference masks from respective datasets. The commercial model omitted structures that were only partially visible or had low confidence. To avoid skewing the plots, data points with DSC of 0 were removed. It should be noted that not all models could segment all structures; hence, not all structure DSC values could be reported.

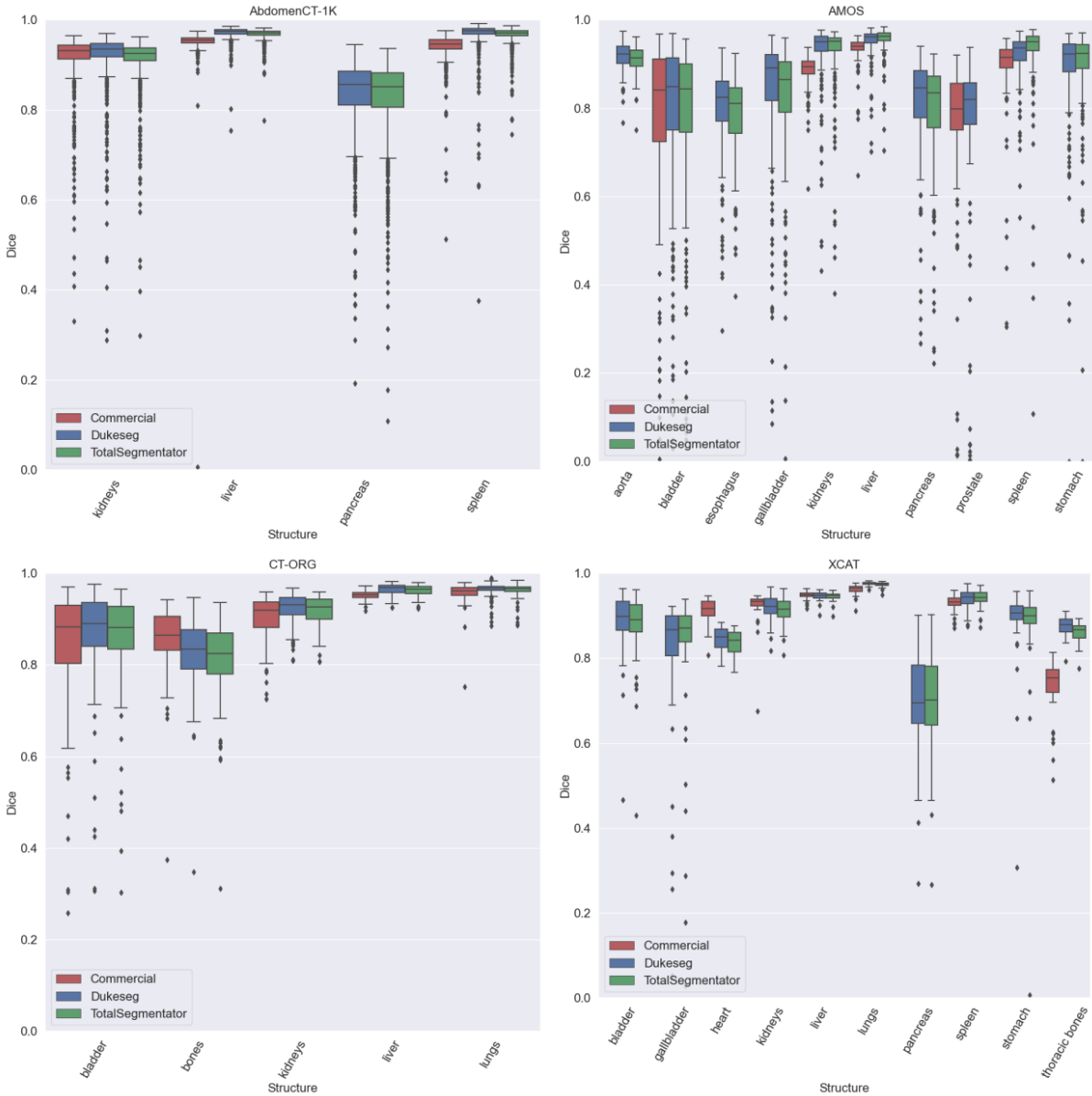

*Figure 3 Boxplots that show the performance of Dukeseg compared with other public and private models. Performance is compared on three public datasets AbdomenCT-1K, AMOS, CT-ORG and a private dataset XCAT. Certain structures like ribs, scapula, and sternum were combined as thoracic bones for the XCAT dataset and minor structures such as the duodenum and adrenal glands were omitted for the AMOS dataset for clarity.*

*3.2 Quality Control*

To evaluate the quality of the phantoms, as noted above, we implemented a multi-step quality control procedure. The first step leveraged bone symmetry by flagging cases where contralateral bones are more than 50% dissimilar in volume. The method proved effective for preliminary screening: 3499 or 98.6% of the 3549 initial cases, met the symmetry volume threshold.

The 3499 volumes from the first quality control step were passed to the volumetric evaluations of different organs. The goal was to exclude patients in which over 25% of the anatomical structures were missing – that is, organ volumes were zero – due to improper or insufficient scanning. This second step yielded 3406 patient CT scans (97.3% from the previous step) that were subsequently converted into phantoms.

Among these 3406 patients, the statistical outlier detection approach provided the outlier probability for every organ of every patient. Notably, the gallbladder exhibited the highest outlier probability, missing in approximately 16% of cases, predominantly due to cholecystectomies. Other structures with frequent anomalies included smaller structures such as the $12^{th}$ ribs, portal and splenic veins, iliac arteries, as well as male-specific organs like the prostate and seminal vesicles. These findings were confirmed by the visual assessments conducted by our physician reviewer.

Approximately 42% of the CT scans were acquired as full body PET/CT scans identified using the statistical clustering method. The filtered patients were not discarded, as there may be false positives, or these still could yield phantoms suitable for many studies. For example, an incomplete skull is irrelevant when creating chest phantoms for cardiothoracic studies. The final step involved filtering patients with multiple CT volumes by selecting those with the lowest outlier probability, resulting in 2,528 unique patients.

*3.3 Patient Demographics and Composition*

In Figure 4, we present the demographic composition of the patient data used to generate the phantoms, reflecting the patient demographics of our institution. The cohort was primarily White, followed by Black, with a smaller representation of Asian individuals and other races categorized as 'Other'. Additionally, the reliance on PET/CT to provide whole body scans introduced a systematic bias, with more male subjects attributable to prostate and lung cancer imaging among those patients. The average age of patients used to create the computational models was $64.9 \pm 14.0$ for males and $61.2 \pm 15.6$ years for females respectively. Furthermore, a heatmap shown on Figure 5 represents the frequency of height and weight within the patient cohort from which the phantoms were created. Approximately ¾ of the patients were concentrated around weight between 70 and 100 kg and height between 1.7 and 1.9 meters. The physical traits that were most common within the patient cohort that contributed to the phantom dataset are shown in this visualization.

In Figure 6, we present the mean organ volume and standard deviation for selected organs across our phantom dataset. Notably, structures such as the gallbladder exhibit a significantly higher standard deviation, reflecting the variability of that small organ and the fact that some patients had undergone gallbladder removal surgery.

In Figure 7, we present a range of phantoms that exhibit a wide variety of age, race, BMI, and sex characteristics to emphasize the diversity within our phantom collection. Muscles and fat have been omitted from these visualizations to maintain an unobstructed view of the internal anatomical features. In the current iteration, bones located below the femur and other smaller bones have not been segmented.

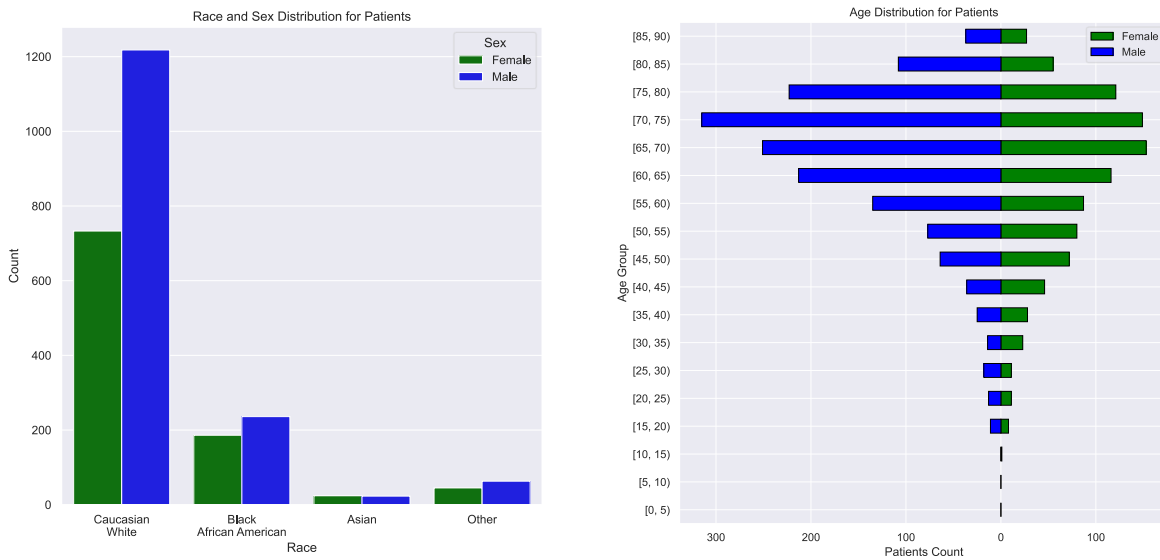

*Figure 4. Race and Age distribution for the 2,528 unique patients used to create the library of phantoms.*

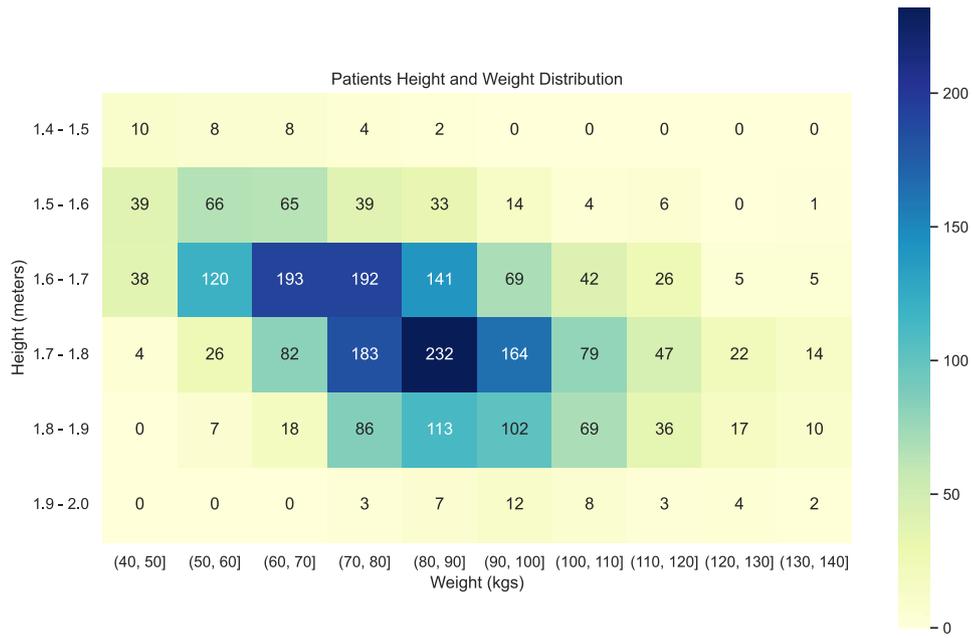

*Figure 5. Height and Weight Distribution of the 2,528 unique patients of the phantoms that passed all quality control measures used to create phantoms.*

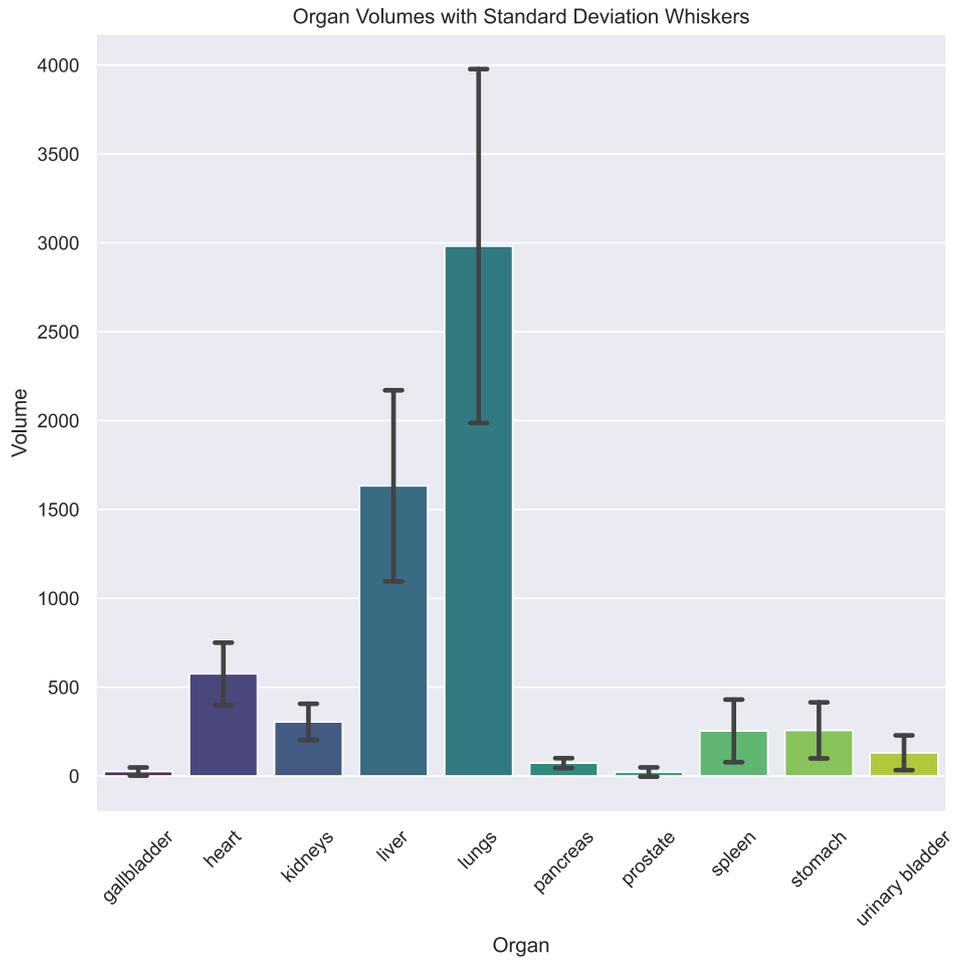

*Figure 6. Organ volumes for selected organs with range of mean and standard deviation. Reflecting diversity of the library, organ volumes can also vary greatly especially for larger organs such as the lungs and liver.*

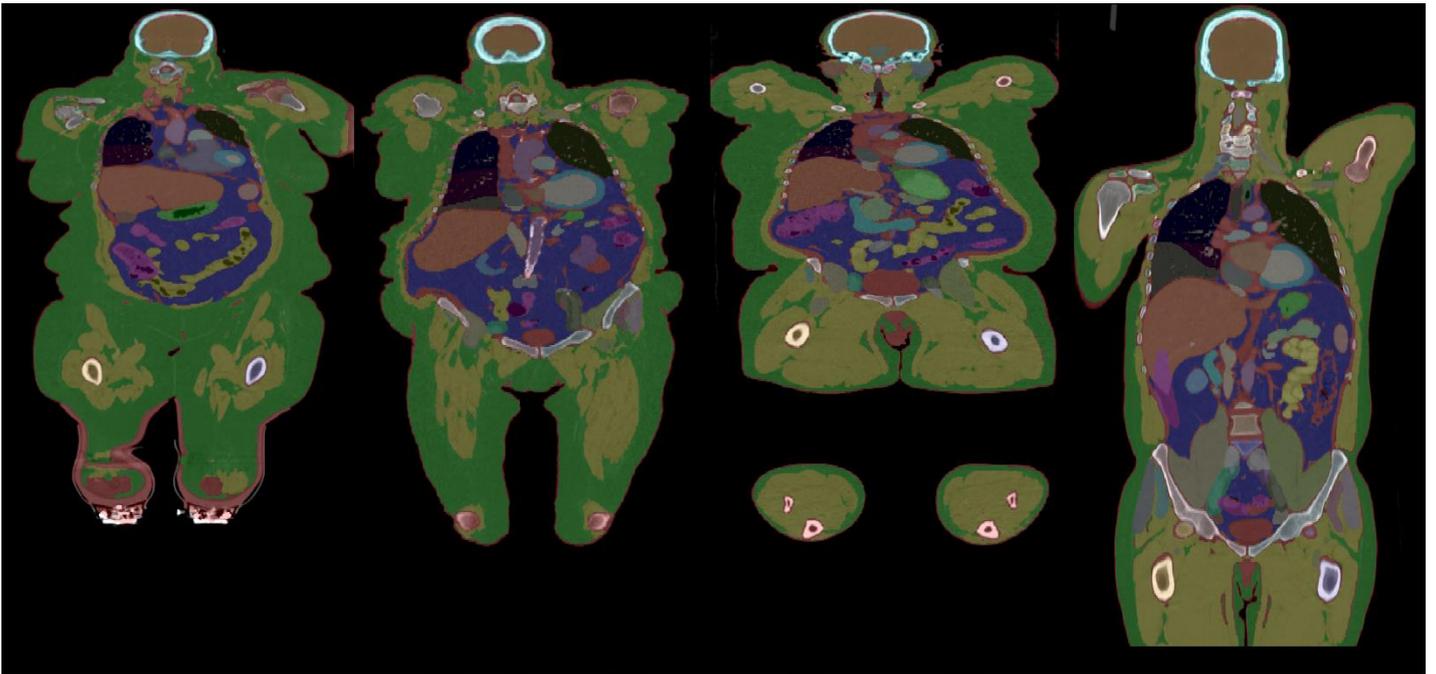

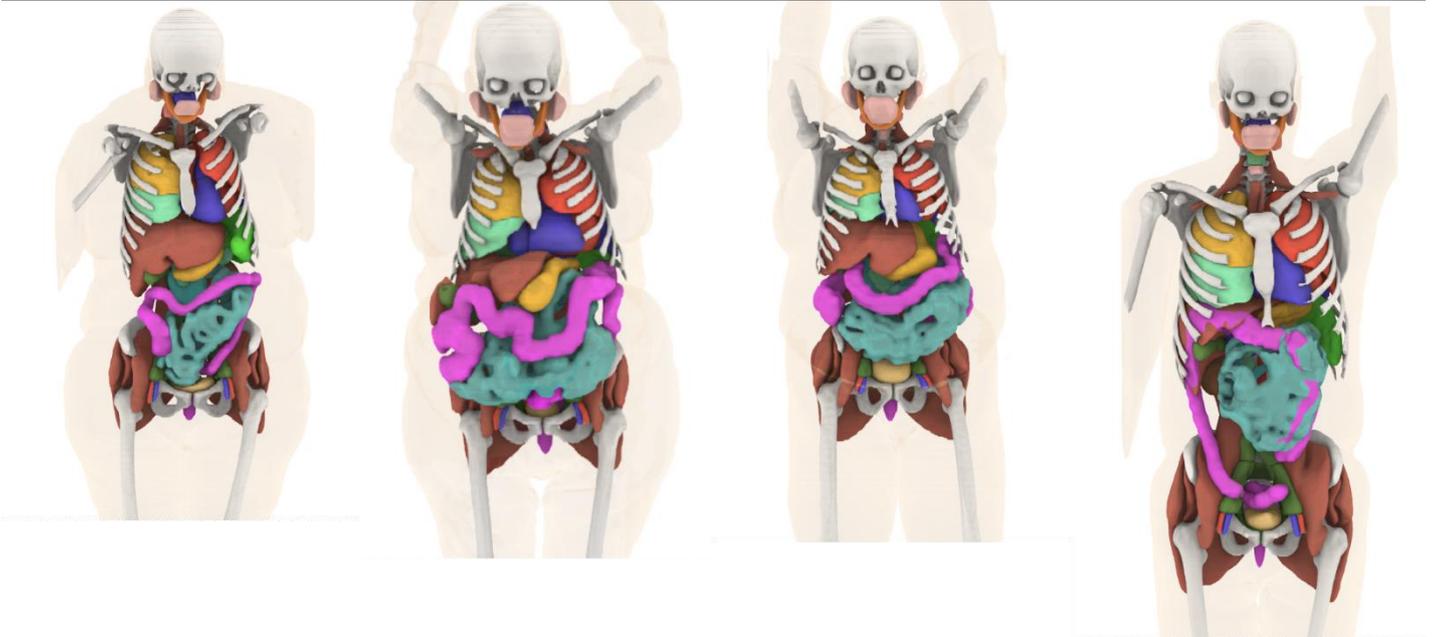

| XCAT2_00277 | XCAT2_02380 | XCAT2_00074 | XCAT2_00294 |
| Sex : Female | Sex : Female | Sex : Male | Sex : Male |
| Age : 69 | Age : 82 | Age : 68 | Age : 44 |
| Height : 1.16 meters | Height : 1.47 meters | Height : 1.75 meters | Height : 1.93 meters |
| Weight : 75 kg | Weight : 73 kg | Weight : 113 kg | Weight : 88 kg |
| Race: White | Race : White | Race : Black | Race : White |

*Figure 7 Qualitative examples: The segmentation overlay on the CT image highlights different structures in various colors. Parrot green indicates subcutaneous fat, yellow represents muscles, red shows structures not segmented by DukeSeg, and other colors are used for the remaining structures (Top); 3D renderings of 4 computational phantoms in increasing order by height.* (Bottom).

# 4 Discussion

In this study, we devised a framework for generating highly detailed, patient-specific anatomical models, which represent human digital twins for research in medical imaging. A significant challenge in the generation of previous phantoms was the manual segmentation of patient CT scans, a prohibitively time-consuming and labor-intensive step. To overcome this bottleneck, we developed a deep learning model capable of segmenting up to 140 structures from patient CT volumes, thus making the phantom generation process substantially more scalable. Furthermore, we implemented an automated quality control module to exclude or flag likely instances of unusual clinical acquisitions or segmentation failure, allowing the new phantoms to be based on the highest quality subset while still preserving the larger library for applications suitable to the end user. This demonstration of scalability in phantom creation was crucial for use in virtual imaging trials, which necessitate a large and diverse sample size to ensure generalizability.

Building a fully automated segmentation pipeline posed significant challenges due to the inherent lack of high-quality image data with validated labels. Even widely used large CT datasets contain incorrect labels, which may limit segmentation model training as well as subsequent external validations. Quality control during the segmentation of 140 structures in this study was a formidable task due to the variations in their size, shape, and appearance. Consequently, a single quality control algorithm proved inadequate. For instance, a quality control measure for small structures like ribs may not be suitable for larger structures like the liver. The development of a multi-step, data-driven quality control process improved the results for this study and may also benefit other segmentation studies.

We have employed a multi-model approach for segmenting the 140 structures, which offers two significant advantages. First, instead of retraining the entire model, we can selectively train only the model of interest. For instance, the skeletal system segmentation model achieved superior results, so we locked down this model early. However, there was still room for improvement in segmenting structures such as the large intestine and small intestine. Since these abdominal structures were segmented using the same model, we were able to focus on enhancing that specific model in further iterations. The second advantage of this design choice pertained to computational efficiency. Our studies involved full body CT scans, typically sized at 512x512x2000 for high resolution CTs at 0.625 mm resolution. Storing output probabilities for 140 classes for each voxel was impractical. Therefore, by employing multiple models, we mitigate the storage challenges associated with inferring and storing probabilities for the entire volume, thereby improving computational efficiency.

The web application developed to display diverse sets of anatomical models allows not only scientific inquiry but also may be helpful as an educational resource, introducing trainees and enthusiasts to detailed human anatomy through a dynamic and engaging platform. This initiative marks a significant stride in enhancing accessibility and interactivity in anatomical education.

# 5 Limitations and Future Work

Our phantoms are derived from CT images, which are obtained from patients undergoing medical evaluations rather than from the general healthy population. This selection bias presents a challenge in representing typical organ volumes accurately. Moreover, the reliance on PET/CT scans to provide whole body coverage led to more male patients. For many applications, the large size of the phantom cohort allows sub-sampling to mitigate these biases. The dataset also included surgical removals such as for the gallbladder or kidney, which can be left unchanged to represent those types of patients or restored generatively or by transforming from an anatomical template. Note, however, that our current patient-specific phantoms faithfully represent the actual anatomy of patient cases undergoing CT imaging.

Although the segmentation model was trained and validated on multiple datasets, the final phantoms were created only using data from Duke University Health System. It is challenging to acquire large clinical datasets that include whole body CT images and radiologist reports to guide the quality control, as well as demographic and clinical data. This restriction means the generated phantoms predominantly represent the demographic distribution of one academic institution in the southeastern US, which may not reflect broader demographics. To address this issue and enhance the model's generalizability, the virtual cohort may be enriched using data from multiple centers. Future studies will expand the scope to improve segmentation techniques and integrate more structures into the computational model by refining the segmentation algorithm or incorporating generative methods to fill in anatomical gaps.

# 6 Conclusion

We developed a pipeline to build computational phantoms using deep learning techniques integrated with a stringent quality control process. We designed an interactive web application with to filter and display phantoms by age, sex, race and BMI. These phantoms will facilitate the conducting of virtual imaging trials in a wide range of clinical applications. This versatility underscores the potential of our phantoms to revolutionize virtual trials and beyond, offering a wide range of possibilities for research and development in the field of medical imaging.

# Data availability

The anthropomorphic phantoms may be requested at [https://cvit.duke.edu/resources/](https://cvit.duke.edu/resources/).

# Code availability

DukeSeg and quality control module is publicly available at https://xcat-3.github.io/. Other post-processing modules can be made available upon request for research purposes.

# Acknowledgments

This work was funded by the Center for Virtual Imaging Trials, NIH grants P41EB028744, R01EB001838, and R01CA261457.